\documentclass[a4paper,twoside,natbib]{article}
\newcommand{\affil}[1]{$^{\rm #1}$}
\usepackage[authoryear]{natbib}
\bibpunct{(}{)}{;}{a}{}{,}
\usepackage{graphicx}
\date{}

\title{\large\bf\flushleft Volume limited dependent Galactic model parameters}
\author{\parbox{\textwidth}{\flushleft
\vspace{-0.5cm}
{\it S. Karaali\affil{A}, S. Bilir\affil{B}, E. Yaz\affil{B}, E. Hamzao\u glu \affil{C} and R. Buser\affil{D}}\\
\vspace{0.4cm}
{\small \affil{A}\, Beykent University, Faculty of Science and Letters, Department of Mathematics and Computer, Beykent 34398, Istanbul, Turkey, Email: skaraali@beykent.edu.tr}\\
{\small \affil{B}\, Istanbul University Science Faculty, Department of Astronomy and Space Sciences, 34119, University-Istanbul, Turkey}\\
{\small \affil{C}\, Beykent University, Faculty of Engineering and Architecture, Department of Computer Engineering, 34398, Istanbul, Turkey}\\
{\small \affil{D}\, Astronomisches Institut der Universit\"{a}t Basel, Venusstrasse 7, 4102 Binningen-Switzerland}}}

\begin{document}
\twocolumn[
\begin{changemargin}{.8cm}{.5cm}
\begin{minipage}{.9\textwidth}
\vspace{-1cm}
\maketitle
\small{\bf Abstract:}
We estimated 34 sets of Galactic model parameters for three intermediate latitude fields with Galactic longitudes $l=60^\circ$, $l=90^\circ$, and $l =180^\circ$, and we discussed their dependence on the volume. Also, we confirmed the variation of these parameters with absolute magnitude and Galactic longitude. The star samples in two fields are restricted with bright and unit absolute magnitude intervals, $4<M_{g}\leq5$, and $5<M_{g}\leq6$, whereas for the third field ($l=60^\circ$) a larger absolute magnitude interval is adopted, $4<M_{g}\leq10$. The limiting apparent magnitudes of star samples are $g_{0}=15$ and $g_{0}=22.5$ mag which provide space densities within distances in the line of sight $\sim$0.9 and 25 kpc. 

The Galactic model parameters for the thin disc are not volume dependent. However, the ones for thick disc and halo do show spectacular trends in their variations with volume, except for the scalelength of the thick disc. The local space density of the thick disc increases, whereas the scaleheight of the same Galactic component decreases monotonically. However, both model parameters approach asymptotic values at large distances. 

The relative local space density of the halo estimated by fitting the density laws to the space densities evaluated for all volumes is constant, except for the small ones. However it is absolute magnitude and Galactic longitude dependent. The axial ratio of the halo increases abruptly for the volumes where thick disc is dominant, whereas it approaches an asymptotic value gradually for larger volumes, indicating a continuous transition from disc–-like structure to a spherical one at the outermost region of the Galaxy. The variation of the Galactic model parameters with absolute magnitude can be explained by their dependence on the stellar luminosity, whereas the variation with volume and Galactic longitude at short distances is a bias in analysis.
      
\medskip
\medskip
\end{minipage}
\end{changemargin}
]
\small

\section{Introduction}
The traditional star-count analyses of the Galactic structure have provided a picture of the basic structural and stellar populations of the Galaxy. Examples and reviews of these analyses can be found in \citet{Bahcall86}, \citet{GWK89}, \citet{Majewski93}, \citet*{Robin00}, and recently \citet{C01} and \citet{Siegel02}. The largest of the observational studies prior the Sloan Digital Sky Survey ({\em SDSS}) are based on photographic surveys. The Basle Halo Program \citep{Becker65} has presented the largest systematic photometric survey of the Galaxy \citep{DF87, FK87, FK90, FK91, F89a, F89b, F89c, F89d}. The Basle Halo Program photometry is currently being recalibrated and reanalysed, using an improved calibration of the {\em RGU} photometric system \citep*{BF90, Ak98, Buser98, Buser99, Karatas01, Karaali04, Bilir04}. More recent and future studies are being based on charge-coupled device (CCD) survey data. 

Our knowledge of the structure of the Galaxy, as deduced from star count data with colour information, entered now to the next step of precision with the advent of new surveys such as {\em SDSS}, {\em 2MASS}, {\em DENIS}, {\em UKIDSS}, {\em VST}, {\em CFH/Megacam} and {\em Suprime}. Researchers have used different methods to determine the Galactic model parameters. The results of these works are summarized in Table 1 of \citet*{KBH04}. One can see that there is an improvement for the numerical values of the model parameters. The local space density and the scaleheight of the thick disc can be given as an example. The evaluation of the thick disc have steadily moved towards shorter scaleheights, from 1.45 to 0.65 kpc \citep{GR83,C01} and higher local densities (2-10\%). In many studies the range of values for the parameters is large. For example, \citet{C01} and \citet{Siegel02} give 6.5-13\% and 6-10\%, respectively, for the local space density for the thick disc. However, one expects the most evolved numerical values from these recent works. That is, either the range for this parameter should be small or a single value with a small error should be given for it. It seems that researchers have not been able to choose the most appropriate procedures for this topic. Finally, we quote the work of \citet{J05}. Although it is a recent work, the Galactic model parameters cited in it are close to the ones claimed in the early works of star– counts, i.e. 4\% relative local space density and 1200 pc scaleheight for the thick disc. However, the size of the field and the number of stars, 6500 deg$^2$ and $\sim$ 48 million stars, are rather different than the ones cited above. Additionally, \citet{J05} admit that, fits applied to the entire dataset are significantly uncertain due to the presence of clumps and overdensities.

Large range or different numerical values, estimated by different researchers, for a specific Galactic model parameter may be due to several reasons: 1) The Galactic model parameters are Galactic latitude/longitude dependent. The two works of \cite*{Buser98,Buser99} cited above confirm this suggestion. Although these authors give a mean value for each parameter, there are differences between the values of a given parameter for different fields. Also, it is shown in the works of \citet{Bilir06a, Bilir06b} and \citet{Cabrera07} that the Galactic model parameters are longitude dependent. 2) The Galactic model parameters are absolute magnitude (stellar luminosity) dependent \citep{KBH04, Bilir06c}. Hence, any procedure which excludes this argument give Galactic model arguments with large ranges. 3) Distance determination also plays an important role on the determination of Galactic model parameters. A single colour magnitude diagram used for a population, for example, gives a mean absolute magnitude for stars with the same colour but with different ultraviolet excesses which results in a single distance for the stars in question. Whereas the procedure based on the colour and metallicity of a star gives individual absolute magnitude for each star and results in more reliable distance determination.

In the present study we show that additional constrains are needed to be taken into consideration for the estimation of more reliable Galactic model parameters. We will see that a model parameter is volume limited dependent. That is, it increases or decreases with the volume which covers the star sample. This is an explanation between the difference of numerical values for a specific Galactic model parameter, estimated in different works where different limiting apparent magnitudes adopted. Contrary to the expectation and physical explanation, the relative local space density of the thick disc also varies with the volume. This is an excellent example for the bias in analysis which is also the main topic of this paper. 

We introduced some simplifications, i.e. we disregarded the giants. However, stars in our sample are not brighter than $g_{0}=15$ mag, hence the number of giants should be small, even if they exist. We quote \citet{Bilir07}, where the percentage of giants relative to the dwarfs are presented. Neither corrections for binarity and overlapping nor accounting for contamination by compact extragalactic objects could be made.

In Section 2, data and reductions are presented. The Galactic model parameters and their dependence on volume are given in Section 3. Finally, Section 4 provides a summary and discussion.

\section{Data and reductions}

The data were taken from {\em SDSS} (DR 5) on the WEB\footnote{http://www.sdss.org/dr5/access/index.html} of three intermediate-latitude fields with longitudes $l=90^\circ$ (F1), $l=180^\circ$ (F2), and $l=60^\circ$ (F3). The Galactic latitudes of the fields F1 and F2 are equal, $b=50^\circ$, whereas it is $b=45^\circ$ for F3. Also, there are differences between the limiting apparent magnitudes of the fields. For F1 and F2, $15.5<g_{0}\leq22.5$ whereas for F3, $15<g_{0}\leq22$. The field F3 has the advantage of providing stars with relatively small distances. Data for the fields are given in Table 1. The total absorption $A_{m}$ (m= $u$, $g$, $r$, $i$ and $z$) for each band is taken from the query server of {\em SDSS} DR5. Thus, the de-reddened magnitudes, with subscript 0, are        

\begin{eqnarray}
u_{0}=u-A_{u}, g_{0}=g-A_{g}, r_{0}=r-A_{r},\\\nonumber
i_{0}=i-A_{i}, z_{0}=z-A_{z}.
\end{eqnarray}

\begin{table}
\tiny{
\caption{Data for the fields. N is the number of stars.}
\center
\begin{tabular}{ccccccc}
\hline
Field & $\alpha$ & $\delta$& {\it l} &{\it b} & Size&  N \\
      & (h~~m~~s)  & ($^{o}~~~~'~~~~''$)& (deg) & (deg) & (deg$^2$)&  \\
\hline
F1 &   15 26 21 &  +56 02 34 & 90  & 50 &  25 & 79411 \\
F2 &   09 47 45 &  +41 19 24 & 180 & 50 &  25 & 65196 \\
F3 &   16 21 34 &  +37 30 30 & 60  & 45 &  10 & 43047 \\
\hline
\end{tabular}
}
\end{table}  

All the colours and magnitudes mentioned hereafter will be de-reddened. Given that the location of the vast majority of our targets are at distances larger than 0.4 kpc, it seems appropriate to apply the full extinction from the maps.

According to \citet{C01}, the distribution of stars in an apparent magnitude--colour diagram, $g_{0}-(g-r)_{0}$, can be classified as follows. The blue stars in the range $15<g_{0}<18$ are dominated by thick disc stars with turn-off at $(g-r)_{0}\sim0.33$, and for $g_{0}>18$ the Galactic halo stars, with turn-off at $(g-r)_{0}\sim0.2$, become significant. Red stars, $(g-r)_{0}\sim1.3$, are dominated by thin disc stars at all apparent magnitudes. 

However, the apparent magnitude-colour diagram and two-colour diagrams for all objects (due to shortage of space large amount of data not presented here) indicate that the stellar distributions are contaminated by extra-galactic objects as claimed by \citet{C01}. Distinction between star/galaxy was obtained using command probPSF given in DR5 WEB page. There 1 or 0 is designated for the probability of objects being a star or galaxy. Needless to say, separation of 1 or 0 strongly depends on seeing and sky brightness. We also applied the ``locus--projection'' method of \citet{J05} in order to remove hot white dwarfs, low-redshift quasars, and white/red dwarf unresolved binaries from our sample. This procedure consists of rejecting objects at distances larger than 0.3 mag from the stellar locus (Figure 1). The apparent magnitude histogram for all objects and for stars is given in Figure 2.  

\begin{figure}
\begin{center}
\includegraphics[angle=0, width=70mm, height=70mm]{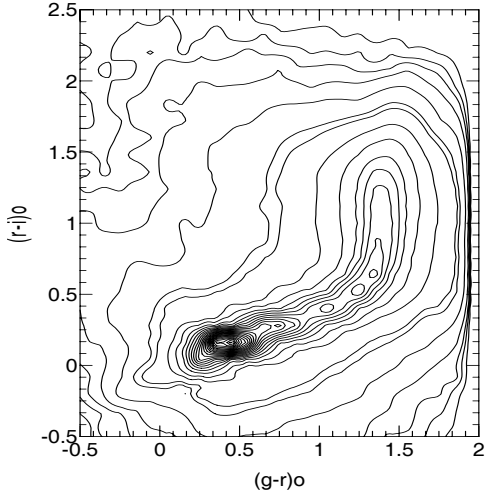}
\caption[]{$(g-r)_{0}-(r-i)_{0}$ two colour diagram for all sources. Stars lie within 0.3 mag of the locus.} 
\label{gr-ri}
\end{center}
\end{figure} 

\begin{figure}
\begin{center}
\includegraphics[angle=0, width=70mm, height=96mm]{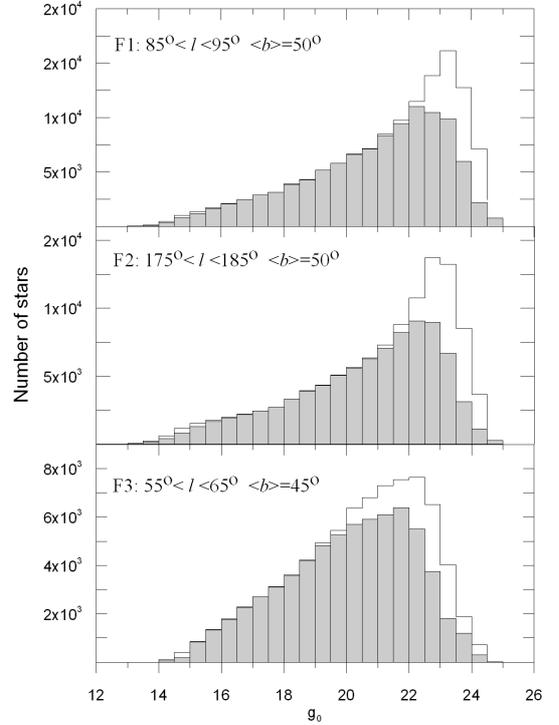}
\caption[]{Apparent magnitude histograms for all sources (white area) and for stars only (shaded area) for the fields F1, F2, and F3.} 
\label{gr-ri}
\end{center}
\end{figure} 

\subsection{Absolute magnitude and distance determination}

We determined two sets of absolute magnitudes, $4<M_{g}\leq5$ and $5<M_{g}\leq6$, for our star sample for the fields F1 and F2, whereas a single but larger set of absolute magnitudes, $4<M_{g}\leq10$, for F3. The halo and thick disc stars are dominant in the absolutely bright intervals, however the relative number of thin disc stars is larger in the interval $4<M_{g}\leq10$. We used the procedure in \citet{KBT05} where the absolute magnitude offset from the Hyades main sequence, $\Delta M^{H}_{g}$, is given as a function of both $(g-r)_{0}$ colour and $\delta_{0.43}$ UV-excess, as follows:

\begin{eqnarray}
\Delta M^{H}_{g} = 
c_{3}\delta^{3}_{0.43}+c_{2}\delta^{2}_{0.43}+c_{1}\delta_{0.43}+c_{0},
\end{eqnarray}
where $\delta_{0.43}$ is the UV-excess standardized to the colour index $(g-r)_{0}=0.43$ in the {\em SDSS} system which corresponds to $\delta_{0.60}$ excess standardized to $(B-V)_{0}=0.60$ in the UBV system; the coefficients $c_{i}$ (i=0, 1, 2, 3) are functions of $(g-r)_{0}$ colour (Table 2) and they are adopted from the work of KBT, and where $\Delta M^{H}_{g}$ is defined as the difference in absolute magnitude of a program star and a Hyades star of the same $(g-r)_{0}$ colour:

\begin{eqnarray}
\Delta M^{H}_{g} = M^{*}_{g}-M^{H}_{g}.
\end{eqnarray}

\begin{table}
\scriptsize{
\center
\caption{Numerical values for the coefficients $c_{i}$ (i=0, 1, 2, 3) in eq. (2).}
\begin{tabular}{crrrr}
\hline
$(g-r)_{o}$ &\multicolumn{1}{c}{$c_{3}$}&\multicolumn{1} {c}
{$c_{2}$} & \multicolumn{1} {c} {$c_{1}$}& \multicolumn{1}
{c} {$c_{0}$}\\
\hline
(0.12,0.22] &   -68.1210 &    26.2746 &     2.2277 &    -0.0177 \\
(0.22,0.32] &   -32.5618 &     6.1310 &     5.7587 &     0.0022 \\
(0.32,0.43] &     8.2789 &    -7.9259 &     6.9140 &     0.0134 \\
(0.43,0.53] &   -23.6455 &    -0.4971 &     6.4561 &     0.0153 \\
(0.53,0.64] &     0.2221 &    -5.9610 &     5.9316 &    -0.0144 \\
(0.64,0.74] &   -47.7038 &     0.1828 &     4.4258 &    -0.0203 \\
(0.74,0.85] &   -52.8605 &    12.0213 &     2.6025 &     0.0051 \\
(0.85,0.95] &   -15.6712 &     7.0498 &     1.6227 &    -0.0047 \\
\hline
\end{tabular}
}
\end{table}
The absolute magnitude for a Hyades star can be evaluated from the Hyades sequence, normalized by KBT (their equation 15). This procedure which is used in our previous works \citep{Ak07a, Ak07b, Bilir07} has two advantages: 1) there is no need to separate the stars into different populations, and 2) the absolute magnitude of a star is determined by its UV-excess individually which provides more accurate absolute magnitudes relative to the procedure in situ where a specific colour magnitude diagram is used for all stars of the same population. When one uses the last two equations (2), (3) and the following one (4) which provides absolute magnitudes for the Hyades stars it gets the absolute magnitude $M^{*}_{g}$ of a star:

\begin{eqnarray}
M^{H}_{g} = -2.0987(g-r)^{2}-0.0008(u-g)^{2}\nonumber\\
                +0.0842(g-r)(u-g)+7.7557(g-r)\nonumber\\
                -0.1556(u-g)+1.9714.
\end{eqnarray}

In a canonical magnitude-limited volume, the distance to which intrinsically bright stars are visible is larger than the distance to which intrinsically faint stars are visible. The effect of this is that brighter stars are statistically overrepresented and the derived absolute magnitudes are too faint. This effect, known as Malmquist bias \citep{M20}, was formalized into the general formula:

\begin{eqnarray}
M_{g}=M_{0}-\sigma^{2}{d\log A(g) \over dg},
\end{eqnarray}
where $M_{g}$ is the assumed absolute magnitude, $M_{0}$ is the absolute magnitude calculated for any star using the KBT calibration, $\sigma$ is the dispersion of the KBT calibration, and $A(g)$ is the counts evaluated at the apparent magnitude $g_{0}$ of any star. The dispersion in absolute magnitude calibration is about 0.25 mag which produces a correction of less than 0.07 mag due to Malmquist bias. This correction was applied to the {\em SDSS} data in our work.

Combination of the absolute magnitude $M_{g}$ and the apparent magnitude $g_{0}$ of a star gives its distance $r$ relative to the Sun, i.e.

\begin{eqnarray}
[g-M_{g}]_{0}=5\log r-5.
\end{eqnarray}

\subsection{Density functions}
The density functions were evaluated for 34 volumes in order to estimate a set of Galactic model parameters for each sample of stars with absolute magnitudes $4<M_{g}\leq5$ and $5<M_{g}\leq6$ for the fields F1 and F2, and with absolute magnitudes $4<M_{g}\leq10$ for the field F3. The lower and upper limiting distances of the volumes are defined such that to obtain reliable model parameters for three Galactic components, i.e. thin and thick discs and halo. The brighter absolute magnitude interval $4<M_{g}\leq5$, provides larger distances. Hence, densities for seven volumes could be evaluated for each field. Whereas for the fainter absolute magnitude interval, $5<M_{g}\leq6$, only six volumes were available. The large absolute magnitude interval $4<M_{g}\leq10$ provides both short and large distances which results densities with eight volumes. There are some differences between the lower and upper limiting distances of the volumes for stars with different absolute magnitudes, i.e. (1.5, 3], (1.5, 5], (1.5, 7.5], (1.5, 10], (1.5, 15], (1.5, 20], and (1.5, 25] kpc for $4<M_{g}\leq5$; (1.25, 2], (1.25, 3], (1.25, 5], (1.25, 7.5], (1.25, 10], and (1.5, 15] kpc for $5<M_{g}\leq6$; and (0.9, 1.5], (0.9, 3], (0.9, 5], (0.9, 7.5], (0.9, 10], (0.9, 15], (0.9, 20], and (0.9, 25] kpc for $4<M_{g}\leq10$. Yet, here we have not provided the corresponding tables due to shortage of space. But all density functions are presented in Figure 3, as $D^{*}=\log D+10$, where $D=N/\Delta V_{1,2}$; $\Delta V_{1,2}=(\pi/180)^{2}(A/3)(r_{2}^3-r_{1}^3)$; $A$ denotes the size of the field; $r_{1}$ and $r_{2}$ denote the lower and upper limiting distance of the volume $\Delta V_{1,2}$; $N$ is the number of stars per unit absolute magnitude; $r^{*}=[(r^{3}_{1}+r^{3}_{2})/2]^{1/3}$ is the centroid distance of the volume $\Delta V_{1,2}$; $z^{*}=r^{*}\sin(b)$, $b$ being the Galactic latitude of the field center. 

\begin{figure*}
\begin{center}
\includegraphics[scale=.85, angle=0]{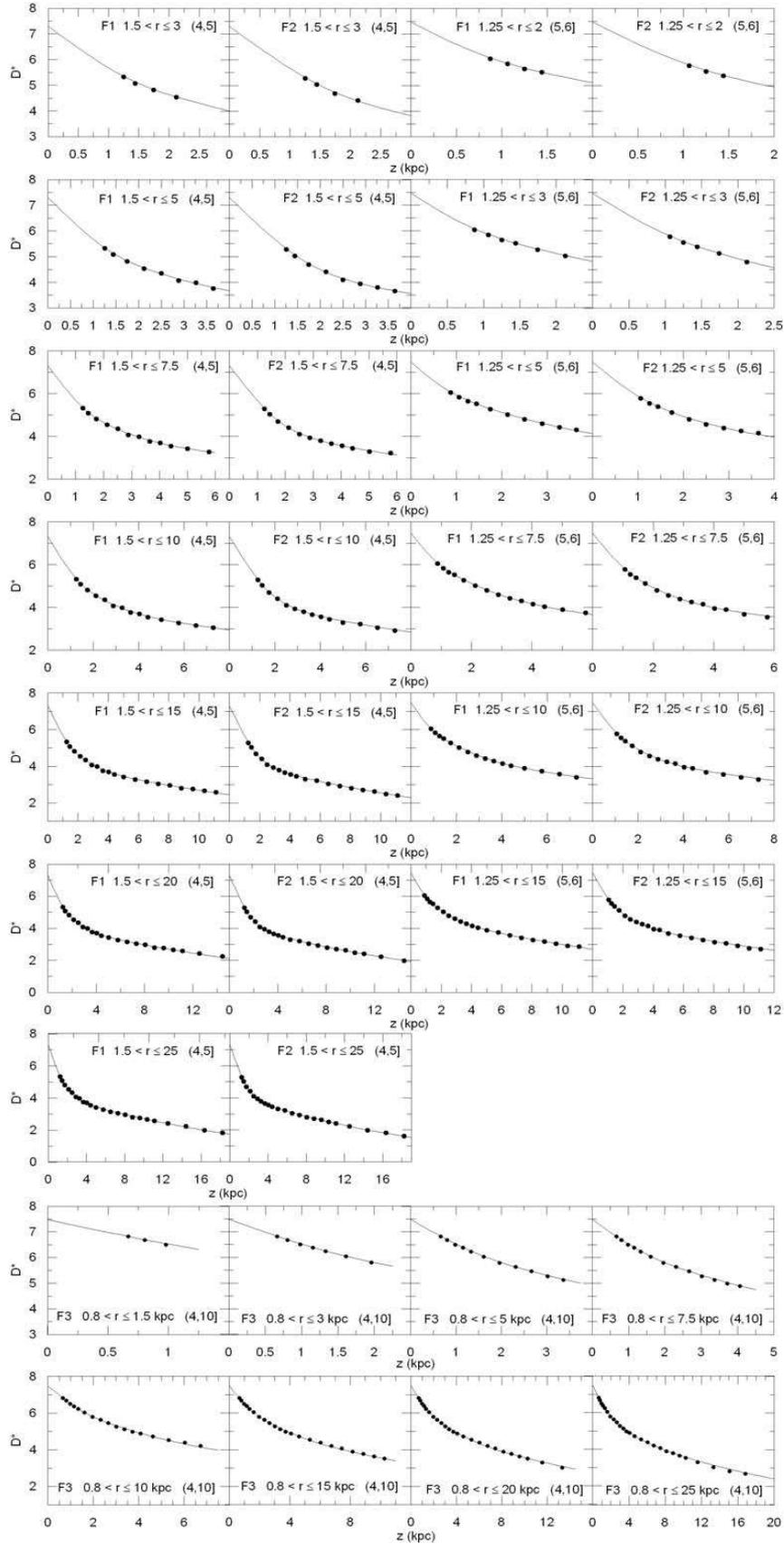}
\caption[]{Comparison of the space density functions evaluated for 34 volumes for the absolute magnitude intervals $4<M_{g}\leq5$ and $5<M_{g}\leq6$ for the fields F1 ($l=90^\circ$) and F2 ($l=180^\circ$), and for $4<M_{g}\leq10$ for the field F3 ($l=60^\circ$).}
\label{gr-ri}
\end{center}
\end{figure*} 

\subsection{Density laws}
In this work we adopted the density laws of Basle Group \citep{Buser98, Buser99}. Disc structures are usually parametrized in cylindrical coordinates by radial and vertical exponentials:
\begin{equation}
D_{i}(R,z)=n_{i}~\exp(-|z|/h_{z,i})~\exp(-(x-R_{0})/h_{i}),\\
\label{ec1}
\end{equation}
where $z=z_{\odot}+r\sin(b)$, $r$ is the distance to the object from the Sun, $b$ the Galactic latitude, $z_{\odot}$  the vertical distance of the Sun from the Galactic plane, 24 pc \citep{J05}, $x$ the projection of the galactocentric distance on the Galactic plane, $R_{0}$ the solar distance from the Galactic centre \citep[8 kpc,][]{R93}, $h_{z,i}$ and $h_{i}$ are the scaleheight and scalelength, respectively, and $n_{i}$ is the normalized density at the solar radius. The suffix $i$ takes the values 1 and 2 as long as the thin and thick discs are considered.

The density law form for the spheroid component used in our work is as follows.
\begin{eqnarray}
D_{s}(R)=n_{s}~\exp[10.093(1-(R/R_{0})^{1/4})]/(R/R_{0})^{7/8},
\end{eqnarray}
where $R$ is the (uncorrected) galactocentric distance in spherical coordinates, and $n_{s}$ the normalized local density. $R$ has to be corrected for the axial ratio $(c/a)$, 
\begin{eqnarray}
R = [x^{2}+(z/(c/a))^2]^{1/2},
\end{eqnarray}
where
\begin{eqnarray}
z = r \sin b,
\end{eqnarray}
\begin{eqnarray}
x = [R_{0}^{2}+r^{2}\cos^{2} b-2R_{0}r\cos b \cos l]^{1/2}, 
\end{eqnarray}
with $r$ the distance along the line of sight, and ($l$,$b$) the Galactic coordinates for the field under investigation.

\section{Galactic model parameters}
We estimated the local space densities and scaleheights for the thin and thick discs and the local space density and the axial ratio for the halo simultaneously by using an {\em in situ} procedure, i.e. by employing a $\chi^{2}$ method to fit the space density functions derived from the observations (combined for all three population components) with a corresponding combination of the adopted population-specific, analytical density laws. For each parameter, we determined its $\chi^{2}$ values by allowing the parameter to vary within its assigned range while keeping all other parameters fixed at their values adopted by the appropriate lowest $\chi^{2}$ model. The range of each parameter and the steps used in their estimation are given in Table 3. Thus, we have 13 sets of Galactic model parameters for thin and thick discs and for the halo, for each of the fields F1 and F2; and eight sets for F3 (Table 4). The sample stars with absolute magnitudes $4<M_{g}\leq5$ and $5<M_{g}\leq6$ are not as close as the stars with $4<M_{g}\leq10$. Hence, the distance interval where the space densities are extrapolated to zero distance is a bit larger. However, they confirm the solar space densities of {\em Hipparcos} \citep{Jahreiss97} and they exhibit similar trend of stars with $4<M_{g}\leq10$ (see the following sections). Hence, the model parameters of absolutely bright stars are as significant as the stars closer to the Sun.      

\begin{table}
\scriptsize{
\center
\caption{The ranges of density law parameters. The symbol $n_{1}$ denotes the local space density of the thin disc.}
\begin{tabular}{ccccc}
\hline
Component   & Parameter     & Unit & Range & Step\\
\hline
Thin disc   & scaleheight   &     pc   &  50--450   &    1 \\
            & scalelength   &    kpc   &   1--3     & 0.025\\
Thick disc  & local density &  $n_{1}$ &  0--25~\%  & 0.01 \\
            & scaleheight   &    pc    &  350--1500 &    1 \\
            & scalelength   &    kpc   &   2--5     & 0.025\\
Halo        & local density &  $n_{1}$ &  0--1~\%   & 0.01 \\
            & flattening   &    --     &  0.1--1.0  & 0.01 \\
\hline
\end{tabular}
}
\end{table}

\begin{table*}
\tiny{
\caption{13 sets of Galactic model parameters for the fields, Field 1 and Field 2, and 8 sets for the field F3, as a function of absolute magnitude and volume. The columns indicate: $M_{g}$– absolute  magnitude interval, lower and upper limiting distance of the volume $r_{1}-–r_{2}$ (kpc), logarithmic local space density for the thin disc $n^{*}_{1}$, scaleheight for the thin disc $H_{1}$, scalelength for the thin disc $h_{1}$, logarithmic local space density for the thick disc $n^{*}_{2}$, scaleheight for the thick disc $H_{2}$, scalelength for the thick disc $h_{2}$, relative local space density of the thick disc $n_{2}/n_{1}$ (\%), logarithmic local space density for the halo $n_{3}^{*}$, the axis ratio of the halo ($c/a$), relative local space density of the halo $n_{3}/n_{1}$ (\%), and $\chi^{2}_{min}$. Scaleheights in pc, scalelengths in kpc.} 
\begin{tabular}{clccccccccccr}
\hline
$M_{g}$& $r_{1}-r_{2}$ & $n^*_{1}$ & $H_{1}$   &  $h_{1}$   & $n^*_{2}$ & $H_{2}$    & $h_{2}$       & $n_{2}/n_{1}$ & $n^*_{3}$ & ($c/a$) & $n_{3}/n_{1}$& $\chi^{2}_{min}$ \\
\hline
Field 1&    &  & & &  & & & & & & & \\
(4, 5]&1.5-3 & 7.287 & 247$\pm$4 & 2.15$\pm$1.00 & 5.756 & 740$\pm$20 & 3.75$\pm$0.75 & 2.94$\pm$0.18 & 3.287 & 0.52$\pm$0.23 & 0.01$\pm$0.01 & 1.19 \\
  & 1.5-5   & 7.287 & 247$\pm$5 & 2.25$\pm$0.85 & 5.763 & 709$\pm$22 & 3.75$\pm$0.95 & 2.99$\pm$0.24 & 3.889 & 0.73$\pm$0.20 & 0.04$\pm$0.02 & 7.57 \\
  & 1.5-7.5 & 7.287 & 247$\pm$5 & 2.20$\pm$0.95 & 5.770 & 706$\pm$19 & 3.62$\pm$0.68 & 3.04$\pm$0.24 & 3.889 & 0.79$\pm$0.17 & 0.04$\pm$0.01 & 8.54 \\
  & 1.5-10  & 7.287 & 247$\pm$5 & 2.30$\pm$0.90 & 5.775 & 700$\pm$21 & 3.70$\pm$0.75 & 3.08$\pm$0.22 & 3.889 & 0.81$\pm$0.13 & 0.04$\pm$0.01 & 9.33 \\
  & 1.5-15  & 7.287 & 247$\pm$5 & 2.20$\pm$0.90 & 5.781 & 698$\pm$19 & 3.70$\pm$0.80 & 3.12$\pm$0.24 & 3.889 & 0.83$\pm$0.13 & 0.04$\pm$0.01 & 17.19\\
  & 1.5-20  & 7.287 & 247$\pm$5 & 2.22$\pm$0.85 & 5.787 & 691$\pm$22 & 3.80$\pm$0.82 & 3.16$\pm$0.22 & 3.889 & 0.84$\pm$0.14 & 0.04$\pm$0.01 & 31.44\\
  & 1.5-25  & 7.287 & 247$\pm$5 & 2.20$\pm$0.88 & 5.787 & 691$\pm$22 & 3.80$\pm$0.82 & 3.16$\pm$0.22 & 3.889 & 0.85$\pm$0.14 & 0.04$\pm$0.01 & 95.24\\
(5, 6]&1.25-2&7.431 & 220$\pm$3 & 2.20$\pm$0.75 & 6.407 & 670$\pm$14 & 4.00$\pm$0.85 & 9.46$\pm$0.52 & 3.431 & 0.25$\pm$0.25 & 0.01$\pm$0.01 & 2.36 \\
  & 1.25-3  & 7.430 & 220$\pm$3 & 2.20$\pm$0.75 & 6.409 & 652$\pm$10 & 4.00$\pm$0.85 & 9.59$\pm$0.56 & 4.430 & 0.46$\pm$0.21 & 0.10$\pm$0.05 & 2.18 \\
  & 1.25-5  & 7.431 & 220$\pm$3 & 2.20$\pm$0.72 & 6.414 & 638$\pm$10 & 3.00$\pm$0.85 & 9.62$\pm$0.74 & 4.605 & 0.50$\pm$0.12 & 0.15$\pm$0.04 & 10.19\\
  & 1.25-7.5& 7.431 & 220$\pm$3 & 2.08$\pm$0.85 & 6.420 & 637$\pm$10 & 3.00$\pm$0.80 & 9.75$\pm$0.66 & 4.605 & 0.55$\pm$0.07 & 0.15$\pm$0.03 & 11.81\\
  & 1.25-10 & 7.431 & 220$\pm$4 & 2.25$\pm$0.70 & 6.430 & 624$\pm$10 & 3.00$\pm$0.80 & 9.98$\pm$0.68 & 4.605 & 0.56$\pm$0.06 & 0.15$\pm$0.03 & 8.77 \\
  & 1.25-15 & 7.431 & 220$\pm$4 & 2.25$\pm$0.70 & 6.430 & 624$\pm$10 & 3.00$\pm$0.80 & 9.98$\pm$0.64 & 4.605 & 0.56$\pm$0.05 & 0.15$\pm$0.03 & 13.74\\
Field 2&   &  & & &  & & & & & & & \\
(4, 5]&1.5-3& 7.288 & 275$\pm$3 & 2.25$\pm$0.88 & 5.726 & 692$\pm$23 & 4.00$\pm$0.78 & 2.74$\pm$0.08 & 4.242 & 0.40$\pm$0.10 & 0.09$\pm$0.04 & 2.26 \\
  & 1.5-5   & 7.287 & 275$\pm$3 & 2.40$\pm$0.80 & 5.749 & 650$\pm$20 & 3.40$\pm$0.82 & 2.90$\pm$0.10 & 4.366 & 0.53$\pm$0.12 & 0.12$\pm$0.02 & 3.08 \\
  & 1.5-7.5 & 7.287 & 275$\pm$3 & 2.45$\pm$0.75 & 5.754 & 638$\pm$20 & 3.80$\pm$0.75 & 2.93$\pm$0.14 & 4.366 & 0.54$\pm$0.08 & 0.12$\pm$0.02 & 5.88 \\
  & 1.5-10  & 7.287 & 275$\pm$3 & 2.50$\pm$0.78 & 5.767 & 635$\pm$22 & 3.22$\pm$0.80 & 3.02$\pm$0.18 & 4.366 & 0.58$\pm$0.06 & 0.12$\pm$0.02 & 10.65\\
  & 1.5-15  & 7.287 & 275$\pm$3 & 2.50$\pm$0.75 & 5.780 & 632$\pm$24 & 3.05$\pm$0.95 & 3.11$\pm$0.24 & 4.366 & 0.58$\pm$0.06 & 0.12$\pm$0.02 & 12.37\\
  & 1.5-20  & 7.287 & 275$\pm$3 & 2.42$\pm$0.80 & 5.791 & 623$\pm$21 & 3.35$\pm$0.75 & 3.19$\pm$0.16 & 4.366 & 0.58$\pm$0.05 & 0.12$\pm$0.02 & 38.07\\
  & 1.5-25  & 7.287 & 275$\pm$3 & 2.40$\pm$0.82 & 5.795 & 621$\pm$21 & 3.40$\pm$0.75 & 3.22$\pm$0.18 & 4.366 & 0.58$\pm$0.06 & 0.12$\pm$0.02 & 42.80\\
(5, 6]&1.25-2&7.443 & 254$\pm$6 & 2.38$\pm$0.62 & 6.246 & 750$\pm$13 & 4.00$\pm$0.95 & 6.35$\pm$0.26 & 3.443 & 0.35$\pm$0.22 & 0.01$\pm$0.01 & 0.67 \\
  & 1.25-3  & 7.440 & 254$\pm$6 & 2.30$\pm$0.65 & 6.286 & 741$\pm$14 & 3.15$\pm$0.78 & 7.01$\pm$0.48 & 3.918 & 0.40$\pm$0.20 & 0.03$\pm$0.02 & 3.50 \\
  & 1.25-5  & 7.437 & 254$\pm$6 & 2.38$\pm$0.75 & 6.318 & 670$\pm$09 & 3.08$\pm$0.92 & 7.60$\pm$0.78 & 4.714 & 0.55$\pm$0.11 & 0.19$\pm$0.07 & 23.59\\
  & 1.25-7.5& 7.437 & 254$\pm$6 & 2.52$\pm$0.68 & 6.329 & 634$\pm$17 & 3.88$\pm$0.52 & 7.80$\pm$0.60 & 4.715 & 0.60$\pm$0.10 & 0.19$\pm$0.05 & 30.59\\
  & 1.25-10 & 7.437 & 254$\pm$6 & 2.52$\pm$0.70 & 6.330 & 634$\pm$17 & 3.85$\pm$0.62 & 7.82$\pm$0.60 & 4.715 & 0.60$\pm$0.10 & 0.19$\pm$0.05 & 42.38\\
  & 1.25-15 & 7.437 & 254$\pm$7 & 2.38$\pm$0.75 & 6.330 & 634$\pm$23 & 4.00$\pm$0.80 & 7.82$\pm$0.78 & 4.680 & 0.60$\pm$0.07 & 0.19$\pm$0.07 & 52.29\\
Field 3&   &  & & &  & & & & & & & \\
(4, 10]&0.9-1.5&7.444 & 360$\pm$10& 1.65$\pm$0.59 & 6.380 &1030$\pm$190& 3.00$\pm$0.85 & 8.63$\pm$0.42 & 4.440 & 0.52$\pm$0.23 & 0.04$\pm$0.61 & 1.21\\
  & 0.9-3&7.444 & 360$\pm$7& 1.70$\pm$0.58 & 6.400 &970$\pm$68& 2.90$\pm$0.95 & 9.04$\pm$1.01 & 4.600 & 0.62$\pm$0.20 & 0.14$\pm$0.25 & 1.81\\
  & 0.9-5&7.444 & 360$\pm$7& 1.70$\pm$0.60 & 6.412 &950$\pm$42& 2.80$\pm$0.90 & 9.29$\pm$0.78 & 4.650 & 0.64$\pm$0.17 & 0.16$\pm$0.13 & 5.66\\
 &0.9-7.5&7.444 & 360$\pm$8& 1.75$\pm$0.58 & 6.419 &935$\pm$40& 2.80$\pm$0.85 & 9.44$\pm$0.86 & 4.700 & 0.67$\pm$0.14 & 0.18$\pm$0.10 & 8.67\\
 & 0.9-10&7.444 & 360$\pm$8& 1.75$\pm$0.62 & 6.420 &925$\pm$42& 2.80$\pm$0.95 & 9.46$\pm$0.94 & 4.750 & 0.70$\pm$0.12 & 0.20$\pm$0.09 &42.62\\
 & 0.9-15&7.444 & 360$\pm$8& 1.75$\pm$0.61 & 6.425 &915$\pm$42& 2.75$\pm$1.00 & 9.57$\pm$0.95 & 4.754 & 0.71$\pm$0.12 & 0.20$\pm$0.09 &49.76\\
 & 0.9-20&7.444 & 360$\pm$7& 1.75$\pm$0.61 & 6.430 &900$\pm$34& 2.45$\pm$0.95 & 9.68$\pm$0.77 & 4.770 & 0.71$\pm$0.12 & 0.21$\pm$0.07 &46.23\\
 & 0.9-25&7.444 & 360$\pm$7& 1.75$\pm$0.61 & 6.440 &880$\pm$32& 2.30$\pm$0.98 & 9.93$\pm$0.76 & 4.780 & 0.73$\pm$0.11 & 0.22$\pm$0.06 &66.09\\
\hline
\end{tabular}  
}
\end{table*}

\subsection{Galactic model parameters of the thin disc}
The logarithmic local space density ($n_{1}^{*}$) of the thin disc is different for different absolute magnitudes, but it is constant for all sets and for three fields, for a given absolute magnitude interval. Numerically, $n_{1}^{*}=7.29$ for the interval $4<M_{g}\leq5$ and $n_{1}^{*}=7.44$ for $5< M_{g}\leq6$ and $4<M_{g}\leq10$. However, the case is different for the scaleheight of the thin disc. It changes with the absolute magnitude and with the direction of the field investigated. For the absolute magnitude interval $4<M_{g}\leq5$, $H_{1}=247$ pc and $H_{1}=275$ pc for the fields F1 and F2, respectively. Whereas for $5<M_{g}\leq6$, $H_{1}=220$ pc and $H_{1}=254$ pc for F1 and F2, respectively. The scaleheight of the thin disc is $H_{1}=360$ pc for the large absolute magnitude interval, $4<M_{g}\leq10$, for the field F3. The scalelength of the thin disc changes with absolute magnitude, with field and with volume, however no trend can be attributed for its variation. The most conspicuous feature for the scalelength of the thin disc is that it is close to 1.7 kpc for the large absolute magnitude interval, whereas it lies within 2.1 and 2.5 kpc for the brighter absolute magnitude intervals.

\subsection{Galactic model parameters of the thick disc}
The Galactic model parameters of the thick disc, i.e. the local space density relative to the local space density of the thin disc ($n_{2}/n_{1}$), the scaleheight ($H_{2}$), and the scalelength ($h_{2}$), vary with the location of the investigated field, with the absolute magnitude, and with the volume which involves the star sample. The last variation could not be observed for the model parameters of the thin disc. We will discuss the trend of the variation for each model parameter in different sections.

\subsubsection{The local space density of the thick disc}
Table 5 shows that the local space density of the thick disc relative to the local space density of the thin disc, ($n_{2}/n_{1}$), is strongly absolute magnitude (stellar luminosity) dependent. A slight dependence on the longitude is also conspicuous. For the field F1 ($l=90^\circ$) $n_{2}/n_{1} \approx3$ and $n_{2}/n_{1}\approx10\%$ for the absolute magnitude intervals $4<M_{g}\leq5$ and $5<M_{g}\leq6$, respectively. For the field F2 ($l=180^\circ)$, $n_{2}/n_{1}$ is rather close to the one for the field F1 for the brighter absolute magnitude interval, $4<M_{g}\leq5$. However, it is less than the one for the fainter absolute magnitude interval, $n_{2}/n_{1}\approx7.5\%$. The local space density of the thick disc estimated for stars with $4<M_{g}\leq10$ for the field F3 ($l=60^\circ$) is $n_{2}/n_{1}\approx9.5\%$, a value which is generally cited in the literature. Additionally, and more important, is that the local space density of the thick disc is volume dependent. That is, it is different in each set in Table 5. Furthermore, there is a good trend in the variation of the local space density of the thick disc for three absolute magnitude intervals and for three fields. It is an increasing function of the centroid distance of the volume involving the star sample and its variation is steeper at short distances. This is more conspicuous for the field F3 where the local space density could be estimated at shorter distances. However, it approaches an asymptotic value at larger distances (Figure 4). 

\begin{figure}
\begin{center}
\includegraphics[scale=0.33, angle=0]{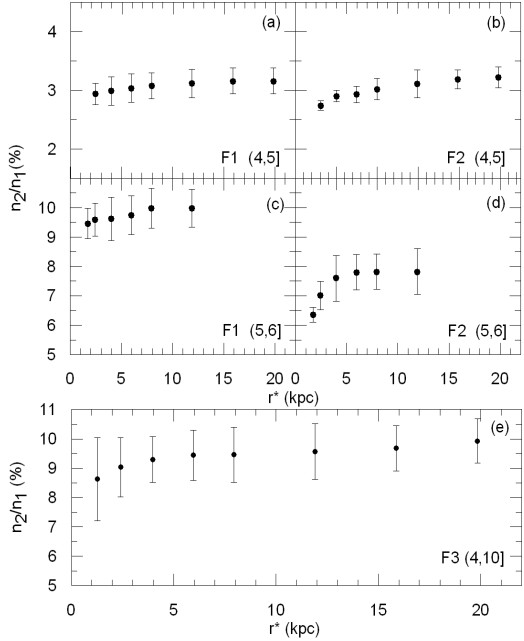}
\caption[]{Relative local space density of the thick disc as a function of absolute magnitude and centroid distance for three fields.} 
\label{gr-ri}
\end{center}
\end{figure} 

\subsubsection{The scaleheight of the thick disc}
As the scaleheight and the local space density are anticorrelated, the scaleheight of the thick disc ($H_{2}$) shows a similar trend as the inferred local space density (Figure 5). However there are some differences between the trends of the scaleheights estimated for a unit absolute magnitude interval, i.e. $4<M_{g}\leq5$ and $5<M_{g}\leq6$, and for the large absolute magnitude interval, $4<M_{g}\leq10$, as explained in the following.  The variation of the scaleheight for stars with $4<M_{g}\leq10$ is steeper at short distances due to the reason explained in the Section 3.2.1. Additionally, the scaleheight for stars with $4<M_{g}\leq5$ and $5<M_{g}\leq6$ is less than 750 pc and it approaches an asymptotic value of $H_{2}\sim650$ pc which is mode value of recent studies. Whereas, for stars with $4<M_{g}\leq10$, the scaleheight of the thick disc lies between 880 and 1030 pc and one can not reveal any asymptotic value from this trend. 

The difference just claimed may be due to the slight difference in Galactic latitude ($\Delta b=5^\circ$) of F3 than the ones of fields F1 and F2, and shorter Galactic longitude ($l=60^\circ$) of the same field (see Section 4 for detail).

\begin{figure}
\begin{center}
\includegraphics[scale=0.40, angle=0]{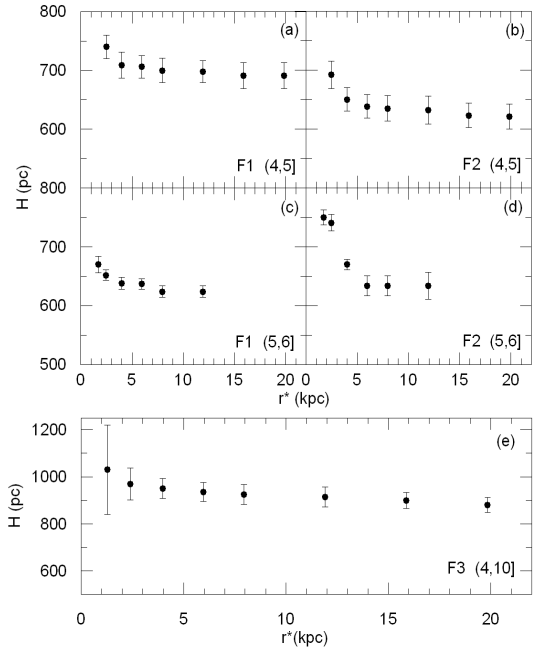}
\caption[]{Scaleheight of the thick disc as a function of absolute magnitude and centroid distance for three fields.} 
\label{gr-ri}
\end{center}
\end{figure} 

\subsubsection{The scalelength of the thick disc}
From Table 4, a constant value can not be attributed to the scalelenght ($h_{2}$) of the thick disc. It varies with volume but the trends are different in five panels (Figure 6). For the sample of stars with absolute magnitudes $4<M_{g}\leq5$, it is flat, $h_{2}\sim3.75$ kpc for the field F1, whereas it is maximum at the least volume, $h_{2}\sim4$ kpc, and minimum at the intermediate volume, $h_{2}\sim3$ kpc for the field F2. For the fainter absolute magnitudes, the variation is a step function at larger distances and it approaches an asymptotic value. The scalelength is $h_{2}=4$ kpc at two small volumes but $h_{2}=3$ kpc at four larger volumes for the field F1, whereas it is almost flat, $h_{2}\sim4$ kpc, for the field F2 except two less values at two intermediate volumes, $h_{2}\sim3$ kpc. The scalelength of the thick disc for the sample of stars with $4<M_{g}\leq10$ is less than or equal to the least scalelength estimated for stars in the fields F1 and F2, i.e. $2.3<h_{2}\leq3$ kpc. The trend is flat at the intermediate distances but it decreases at short and large distances.

\begin{figure}
\begin{center}
\includegraphics[scale=0.33, angle=0]{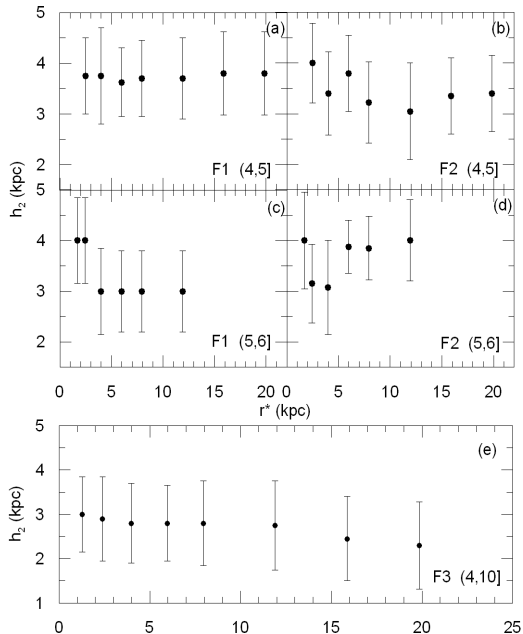}
\caption[]{Scalelength of the thick disc as a function of absolute magnitude and centroid distance for three fields.} 
\end{center}
\end{figure} 

\subsection{The Galactic model parameters of the halo}
The Galactic model parameters of the halo, i.e. the local space density relative to the local space density of the thin disc ($n_{3}/n_{1}$) and the axial ratio ($c/a$), are also luminosity dependent and they are different for star samples at different locations of the Galaxy. The variation of ($n_{3}/n_{1}$) is the same of the variation of the local space density for the thick disc, i.e. it is steeper at short distances but, it approaches an asymptotic value at larger distances. This is more conspicuous for the field F3 where the local space density could be estimated at shorter distances (Table 4). The asymptotic local space densities for the fields F1, F2 and F3 are 0.04$\%$ (for $4<M_{g}\leq5$), 0.15$\%$ (for $5<M_{g}\leq6$); 0.12$\%$ (for $4<M_{g}\leq5$), 0.19$\%$ (for $5<M_{g}\leq6$); and 0.21$\%$ (for $4<M_{g}\leq10$) respectively. The halo is dominant at large distances of our Galaxy. Hence, the asymptotic local space densities just claimed can be attributed as the local space densities for the halo. However, different local space densities for stars with the same absolute magnitudes but at different direction of the Galaxy, such as $n_{3}/n_{1}=0.04\%$ and $n_{3}/n_{1}=0.12\%$ for the fields F1 and F2, has to be explained (see Section 4).
      
The trend of the variation of ($c/a$) for five panels in Figure 7 is the same, i.e. the axial ratio of the halo increases monotonically up to a distance but it becomes flat beyond this distance. However there are some differences between the variations of ($c/a$) for stars with different absolute magnitudes in different fields. For fields F1 and F2 where stars with a unit absolute magnitude interval is considered, one can observe an asymptotic value different from each other, however. Whereas, the variation of ($c/a$) for stars in the field F3 gives the indication that it still increases at larger distances. Different ($c/a$) for different absolute magnitudes show that halo stars of different luminosity occupies different regions within the halo component, and different values for the axial ratio at different distances show that the halo is disc–-like at short distances but spherical at larger distances.
    
\begin{figure}
\begin{center}
\includegraphics[scale=0.33, angle=0]{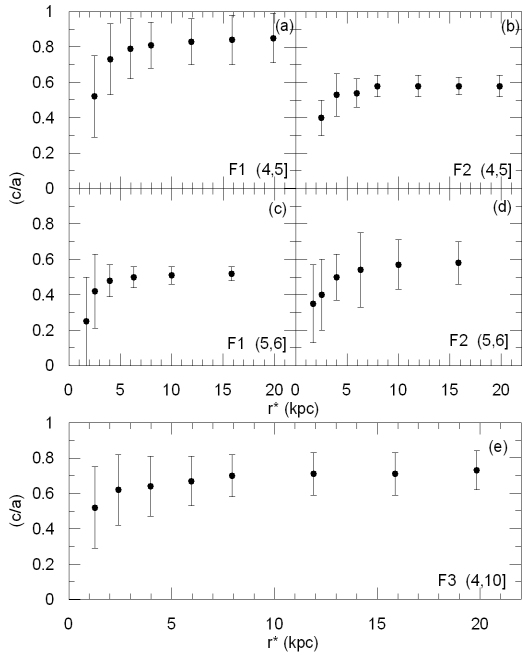}
\caption[]{Axial ratio of the halo as a function of absolute magnitude and the centroid distance for three fields.} 
\end{center}
\end{figure} 

\begin{table}
\scriptsize{
\caption{Relative local space density $n_{2}/n_{1}$(\%) and the scaleheight of the thick disc, $H_{2}$ (pc) for the fields F1, and F2, as a function of absolute magnitude, $M_{g}$, and centroid distance, $r^{*}$. The symbol $r_{1}–-r_{2}$ as in Table 4. Distances are in kpc.}
\begin{tabular}{ccccccc}
\hline
\multicolumn{3} {c} {Field $\rightarrow$} & \multicolumn{2} {c} {F1} &  \multicolumn{2} {c} {F2}\\
\hline
$M_{g}$  & $r_{1}-r_{2}$  & r$^{*}$& $n_{2}/n_{1}$ &    $H$ & $n_{2}/n_{1}$ & $H$ \\
\hline
 (4,5]       &  1.5-3   &  2.48 &  2.94 & 740 & 2.74 & 692 \\
             &  1.5-5   &  4.00 &  2.99 & 709 & 2.90 & 650 \\
             &  1.5-7.5 &  5.97 &  3.04 & 706 & 2.93 & 638 \\
             &  1.5-10  &  7.95 &  3.08 & 700 & 3.02 & 635 \\
             &  1.5-15  & 11.91 &  3.12 & 698 & 3.11 & 632 \\
             &  1.5-20  & 15.88 &  3.16 & 691 & 3.19 & 623 \\
             &  1.5-25  & 19.84 &  3.16 & 691 & 3.22 & 621 \\
 (5,6]       &  1.25-2  & 1.71  &  9.46 & 670 & 6.47 & 750 \\
             &  1.25-3  & 2.44  &  9.59 & 652 & 6.84 & 741 \\
             &  1.25-5  & 3.99  &  9.62 & 638 & 7.01 & 670 \\
             &  1.25-7.5& 5.96  &  9.75 & 637 & 7.50 & 634 \\
             &  1.25-10 & 7.94  &  9.98 & 624 & 7.89 & 634 \\
             &  1.25-15 &11.91  &  9.98 & 624 & 8.13 & 634 \\
\hline
\end{tabular}
}
\end{table}

\subsection{Degeneracy}
No degeneracy could be detected in our work as tried to be explained in the following: 1) The variation of statistics $\chi^2$ for every model parameter exhibit a parabola with a perfect minimum. Figure 8a gives the variation of $\chi^2$ for the scaleheight of thick disc for $4<M_{g}\leq5$, for the field F1, as an example. 2) We used the model parameters of halo estimated via deeper samples which get better constraint on the halo, and omitted the corresponding space density from the total space density of stars with $4<M_{g}\leq5$ for the field F1. We fitted the remaining space density function to a model of two components (thin and thick discs) and we estimated their parameters. They are equal exactly the corresponding ones estimated via fitting the total density functions to a model of three Galactic components (Figure 8b-c). 3) Finally, we changed the counts of stars with $4<M_{g}\leq5$ in F1 by $\pm\sqrt{N}$, adopted the parameters from the field F2 for stars with the same absolute magnitude and we noticed that they do not fit (Figure 8d). The results obtained by the procedures applied to the mentioned sample of stars can be extended to other sample of stars. We conclude that no degeneracy exist in our work.  

\begin{figure}
\begin{center}
\includegraphics[angle=0, width=70mm, height=174mm]{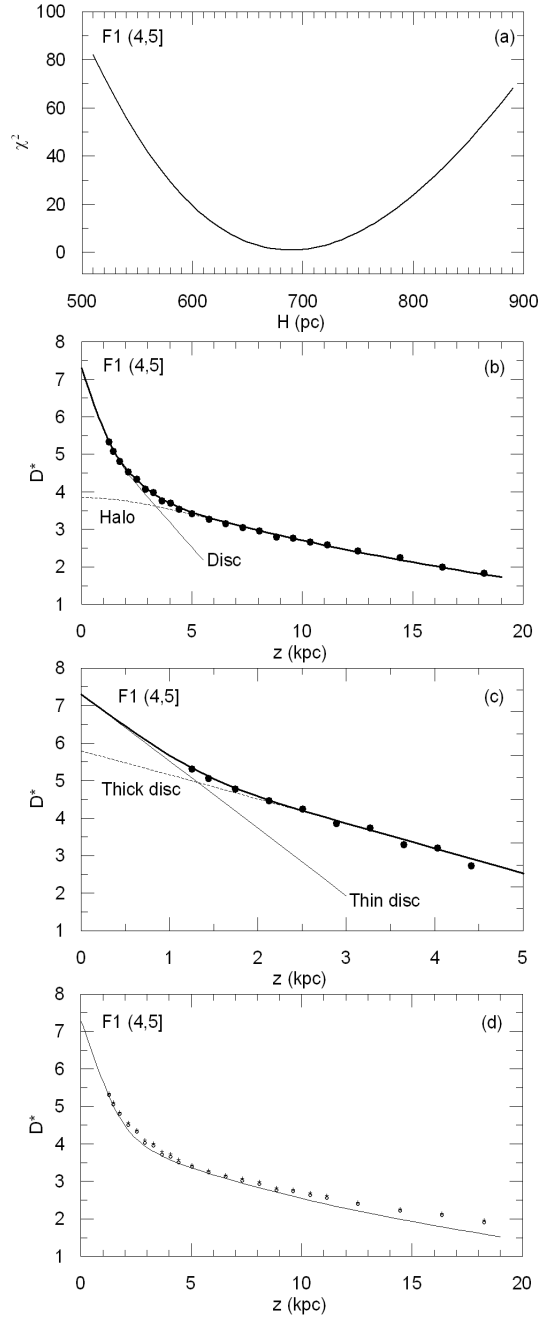}
\caption[]{Four panels related to the discussion of degeneracy. (a) the variation of the $\chi^{2}$ with a perfect minimum, (b) comparison of the logarithmic density function for stars with $4<M_{g}\leq5$ in the largest volume for the field F1 with best fitted Galactic model combined for three populations, thin and thick discs, and halo, (c) comparison of the logarithmic density function for stars in panel (b) with best fitted Galactic model for only thin and thick discs, after omitting the space densities corresponding the model parameters estimated in panel (b) for the halo, (d)  comparison of the logarithmic space density function evaluated by changing the star counts $\pm \sqrt{N}$, for the stars with $4<M_{g}\leq5$ for the field F1, with the Galactic model fitted to the space density function of stars in the field F2. Symbols ($+$) and (o) correspond to star counts $N+\sqrt{N}$ and $N-\sqrt{N}$, respectively.} 
\end{center}
\end{figure}

\section{Summary and Discussion}
We estimated 34 sets of Galactic model parameters for three fields, F1 ($l=90^\circ$), F2 ($l=180^\circ$), and F3 ($l=60^\circ$) and discussed their dependence on absolute magnitude (stellar luminosity), Galactic longitude, and volume. The star samples in F1 and F2 are limited with absolute magnitudes $4<M_{g}\leq5$ and $5<M_{g}\leq6$, whereas the range of the star sample in F3 is adopted larger, $4<M_{g}\leq10$, to provide space densities at shorter distances to the Sun and to compare the resulting Galactic model parameters with the ones in the literature.
     
All Galactic model parameters of the thin and thick discs, and halo are luminosity dependent. That is, a specific model parameter for a component of the Galaxy changes with the considered absolute magnitude interval. The parameters are also Galactic longitude dependent, i.e. a given parameter is different for the fields F1 and F2 for two samples of stars with the same absolute magnitudes. These findings are not new, only they confirm our previous results 
\citep{KBH04, Bilir06c}. The original finding in this work is the effect of the volume. Galactic model parameters varies with the volume of the star sample for a fixed absolute magnitude interval, for a field. However, the trends of the variations of the model parameters with absolute magnitude, location, and volume are different for the Galactic components, i.e. thin and thick discs, and halo.

\subsection{The thin disc}
For the thin disc, the logarithmic local space density ($n^{*}_{1}$) is the same for a field for a given absolute magnitude interval, and it is constant for all volumes in question. Also, the scaleheight of the thin disc is constant for all volumes, however it is different for different fields and different absolute magnitude intervals. The only model parameter of the thin disc which varies with volume, additional to its variation with absolute magnitude and longitude, is the scalelength ($h_{1}$). It lies between 2 and 2.5 kpc, without any trend however. 

\subsection{The thick disc}
The behaviour of the local space density and the scaleheight of the thick disc is different, i.e. they are limiting volume dependent, additional to  absolute magnitude and Galactic longitude. The local space density of the thick disc relative to the local space density of the thin disc ($n_{2}/n_{1}$) increases monotonically at short distances and it approaches an asymptotic value gradually. However, its range is different for different absolute magnitudes and different fields, i.e. for the fields F1 and F2: 2.94-–3.16$\%$ and 2.74-–3.22$\%$ for the absolute magnitude interval $4<M_{g}\leq5$, 9.46-–9.98$\%$ and 6.35–-7.82$\%$ for $5<M_{g}\leq6$, respectively, and 8.63–-9.93$\%$ for the field F3 ($4<M_{g}\leq10$). The relatively small $n_{2}/n_{1}$ values for the absolute magnitude interval $4<M_{g}\leq5$ for the fields F1 and F2 is due to the reason that the thick disc stars are rary in this interval. 

The small values of the local space density remind us the finding of \citet{GR83} who claimed $n_{2}/n_{1}\sim2\%$ for the relative local space density for the thick disc. \citet{GR83} claimed a bright apparent magnitude ($I=18$ mag) and an intermediate distance from the galactic plane in their work which does not contradict with the results of our work. From the other hand, the asymptotic value $n_{2}/n_{1}\sim10\%$ for the fields F3 and F1 ($5<M_{g}\leq6$) is consistent with the work of \citet{Bensbyetal05} who estimated the local normalization of the thick disc as 10$\%$ by kinematical criteria. The value of \citet{J05}, i.e. $n_{2}/n_{1}\sim4\%$, estimated for a data set which covers 6500 deg$^2$ of the sky is close to the  small values in our work. The distance range and the number of stars in their sample are 0.1–-15 kpc, and $\sim$48 million, respectively. However, \citet{J05} admit that fits applied to the entire dataset are significantly uncertain due to the presence of clumps and overdensities which is not the case in our work.

The scaleheight of the thick disc ($H_{2}$) decreases monotonically up to $z\sim4$ kpc, then becomes flat up to $z\sim15$ kpc. For the fields F1 and F2, it approaches an asymptotic value, whereas the scaleheight of the thick disc for the field F3 still decreases, though with a small gradient. The smallest and largest values for $H_{2}$ are estimated for the samples of stars with $4<M_{g}\leq5$ (Field F2) and stars with $4<M_{g}\leq10$ (Field F3), i.e. 621 and 1030 pc, respectively. As mentioned in the Introduction, many authors give a range for the Galactic model parameters (see Table 1 of \cite{KBH04}) . For example \citet{C01} give the range 580–-750 pc which is rather close to the range of $H_{2}$ in our work estimated for the unit absolute intervals for F1 and F2, i.e. 621–-750 pc. From the other hand, the scaleheight of the thick disc claimed by \citet{Du03}, 640 pc, is equal to the mean of the asymptotic scaleheight estimated for the fields F1 and F2, i.e. 621, 624, 634, and 691 pc. The range of the scaleheight (880-–1030 pc) estimated for the large absolute magnitude interval, $4<M_{g}\leq10$, for the field F3 is different than the ones claimed above. One can reveal the indication from Figure 5e that the asymptotic value of the $H_{2}$ is lower than 880 pc. The upper limit of the scaleheight, $H_{2}=1030$ pc, reminds us the values claimed for this parameter before the year 2000. Actually \citet{DF87}, \citet{KG89}, \citet{L96}, and \citet{Buser98, Buser99} give $H_{2}$=1, 1, 0.98, and 0.91 kpc, respectively. However, there are some recent works where large scaleheights were claimed, i.e. the upper limit of \citet{Siegel02}, $H_{2}=1$ kpc, and the most recent work of \citet{J05}, $H_{2}=1.2$ kpc.

\subsection{The halo}
The behaviour of the local space density and the axial ratio of the halo are also limiting volume, absolute magnitude (stellar luminosity) and Galactic longitude dependent. The local space density of the halo relative to the local space density of the thin disc ($n_{3}/n_{1}$) is an increasing function for a short distance range, $z\leq2.5$ kpc for F1 and F2 and $z\leq3.5$ kpc for F3, then it approaches an asymptotic value gradually. Halo stars are dominant at large distances, hence we can consider only the asymptotic local space densities. For the field F1, $n_{3}/n_{1}=0.04\%$ ($4<M_{g}\leq5$) is close to the one claimed by \citet{Buser98, Buser99}, 0.0005; and $n_{3}/n_{1}=0.15\%$ ($5<M_{g}\leq6$) equals exactly to those of \citet{RW93}, \citet{R96, Robin00}, and \citet{Siegel02}. Additionally this value is rather close to the local space density of the halo claimed by many other authors. For the field F2, $n_{3}/n_{1}=0.12\%$ ($4<M_{g}\leq5$) either equals exactly or is close to the local space density of the halo claimed by \citet{TM84}, \citet{YY92}, \citet{C01}, and \citet{Du03}; and $n_{3}/n_{1}=0.19\%$ ($5<M_{g}\leq6$) is very close to corresponding local space densities of \citet{GR83}, \citet{G84}, and \citet{KG89}. We should keep in mind that the cited local space densities for the halo estimated via star count analysis refer to star samples with a large absolute magnitude interval, whereas the ones in our work are estimated for a unit absolute magnitude interval, i.e. $4<M_{g}\leq5$ and $5<M_{g}\leq6$. The relative local space density for the absolute magnitude interval ($4<M_{g}\leq10$) $n_{3}/n_{1}=0.20\%$, is close to the previous one. 

The axial ratio of the halo, ($c/a$), show the same trend in five panels for three fields, however there are some differences between them. It increases within a small distance interval ($z\leq3.5$ kpc for F1 and F2; and $z\leq5$ kpc for F3) and it becomes flat for larger distances. The axial ratio approaches an asymptotic value in panels (a)-–(d), whereas there is an indication in panel (e) of Figure 7 that ($c/a$) still increases. The asymptotic axial ratio for the star samples with $5<M_{g}\leq6$ for the fields F1 and F2, and with $4<M_{g}\leq5$ for the field F2 lie between 0.56 and 0.60 and it is close to the corresponding one cited in recent years \citep[cf.][]{C01, Siegel02, Du03}. Whereas, for stars with $4<M_{g}\leq5$ in the field F1, the asymptotic axial ratio is higher, ($c/a$)=0.85, and it is the mode value cited in early works of star counts (cf. \citet {GR83}, \citet{KG89}. The recent works in which an axial ratio is cited close or equal to the one in question are those of \citet{Buser98, Buser99} and \citet {Ojha99}(their upper limit). Again, we should note that the cited axial ratios are estimated for star samples with large absolute magnitude intervals, whereas our samples are restricted with a unit absolute magnitude interval. Hence, the agreement between our model parameters with the cited ones is interesting. The largest axial ratio for the star sample with $4<M_{g}\leq10$ (field F3) is ($c/a$)=0.73, and it lies in the axial ratio interval claimed by \citep{R96, Robin00} and \citet{Ojha99}.                     

{\bf Conclusion:} The Galactic model parameters estimated for 34 sample of stars in three fields show variation with absolute magnitude, Galactic longitude, and volume. The variation with absolute magnitude can be explained by the dependence of the model parameters on stellar luminosity. Different local space densities for different absolute magnitudes of {\em Hipparcos} \citep{Jahreiss97} is a good confirmation for this argument. However, the case is different for Galactic longitude and volume. The local space density of a population, for example, is fixed due its definition and it must not change with volume or distance. Hence, the only explanation of its variation (in our work) can be done by a bias in the analysis. The gradient of all model parameters is large for small distances because the contributions of three populations, i.e. thin and thick discs, and halo, to the space density change with distance. Whereas, at larger distances the contributions of (thin \& thick) discs diminish, and the gradient approaches zero. Thus the asymptotic value of the corresponding variation can be adopted as the Galactic model parameter in question.

The perfect space density functions in our work reject any possible presence of clumps and overdensities in the fields investigated. Also, the variation of $\chi^2$ for a given model parameter can be fitted to a parabola which provides a perfect minimum, i.e. $\chi^2_{min}$. This is a clue for the absence of any degeneracy in our work. Also, the combination of two fields confirm the non–-degeneracy. 

\section*{Acknowledgments}
Thanks to anonymous referee, for suggestions that have improved the overall quality of the work presented in this
paper. We also thank Hikmet \c{C}akmak for preparing the computer programme for this study. This work was supported by the Research Fund of the University of Istanbul. Project numbers: BYP 832/29112005 and BYP 941/02032006. Two of us (S. Karaali and E. Hamzao\u glu) thank to T.C. Beykent University for financial support.

\end{document}